\documentclass[12pt]{iopart0}
\RequirePackage{iopams0}
\RequirePackage{setstack0}
\usepackage{setstack0}
\usepackage{iopams0}
\renewcommand{\d}[2]{{#1}_{\mathrm{#2}}}
\renewcommand{\u}[2]{{#1}^{\mathrm{#2}}}
\newcommand{\Lie}[2]{\underset{#1}{\mathcal{L}} {#2}}

\begin{document}
\title[Lagrangian structures, integrability and chaos for 3D dynamical equations]{Lagrangian structures, integrability and chaos for 3D dynamical equations}%
\author{Miguel D Bustamante\addtocounter{footnote}{-1}${}^{a}$ and Sergio A Hojman${}^{b,\, a,\, c}$}
  \address{${}^a$ Departamento de F\'{\i}sica, Facultad de Ciencias,
Universidad de Chile,\\ Casilla 653, Santiago, Chile}
 \address{${}^b$
Centro de Recursos Educativos Avanzados, CREA,\\ Vicente P\'erez Rosales
1356-A, Santiago, Chile }
 \address{${}^c$ Facultad de Educaci\'on,
Universidad Nacional Andr\'es Bello,\\ Fern\'andez Concha 700, Santiago,
Chile}
\begin{abstract}
In this paper we consider the general setting for constructing Action
Principles for three--dimensional first order autonomous equations. We
present the results for some integrable and non--integrable cases of the
Lotka--Volterra equation, and we show Lagrangian descriptions which are
valid for systems satisfying Shil'nikov criteria on the existence of strange
attractors, though chaotic behavior or homiclinic orbits have not been
verified up to now. The Euler--Lagrange equations we get for these systems
usually present ``time reparameterization" symmetry, though other kinds of
invariance may be found according to the kernel of the associated symplectic
$2$--form. The formulation of a Hamiltonian structure (Poisson brackets and
Hamiltonians) for these systems from the Lagrangian viewpoint leads to a
method of finding new constants of the motion starting from known ones,
which is applied to some systems found in the literature known to possess a
constant of the motion, to find the other and thus showing their
integrability. In particular, we show that the so--called $ABC$ system is
completely integrable if it possesses one constant of the motion.
\\

\noindent \textbf{Pacs numbers:}  45.20.Jj, 02.30.Ik, 05.45.Ac\,\\

\noindent \textbf{Keywords:} Lagrangian, integrable system, chaos,
Lotka--Volterra\,
\end{abstract}
\eads{\mailto{miguelb@macul.ciencias.uchile.cl},\,\,\mailto{shojman@creavirtual.org}}
\maketitle
\section{Introduction}

Three--dimensional autonomous first order nonlinear systems show a very rich
behavior, from complete integrability\cite{Str88,Cai00,Cai00a,{Mou01}} to
chaos and strange attractors\cite{Van78,Arn80,Arn81,Ros76,{Phi00}},
according to the values of the parameters that appear in the equations of
motion. As they usually represent physical and biological systems of
interest (Lorenz equation for hydrodynamic flow, R\"ossler equation for
chemical reactive systems, Lotka--Volterra equation for laser physics,
 plasma physics,\cite{Lav75} biology,\cite{May75,Abr98}
economy,\cite{Sol99} etc.), it is somewhat surprising the lack of Action
Principles for dealing with these equations. We intend to fill this gap
using the general setting for Lagrangian structures,\cite{Hoj81} which when
applied to three--dimensional systems reduces to a simpler problem, namely:
for a given three--dimensional vector field $V\u{}{a}$, to find the
determinant of a metric (or {volume element}) such that the covariant
divergence $\nabla\d{}{a} V\u{}{a} = 0 $, which is a way of writing
Liouville's theorem and also a condition which is known to lead to the
construction of constants of the motion starting from symmetries
only\cite{Hoj92}. The Euler--Lagrange equations that come out from these
principles are usually equivalent to $\frac{d x\u{}{a}}{d s} = V\u{}{a}$,
where $s$ is an arbitrary parameter (time reparameterization symmetry).

In section \ref{secPreview} a preview of known and new results on the
Lagrangian approach for three--dimensional systems is presented. Section
\ref{secApp} deals with applications of the theory: a method to construct
new constants of the motion from known ones, construction of Hamiltonian
structures, and several kinds of Action Principles with invariance related
to symmetries. In section \ref{secExamples} we show examples for the
Lotka--Volterra equation, including integrable cases as well as
non--integrable ones: we show that recently found ``quasi--integrable"
cases\cite{Cai00a} (i.e., systems possessing one constant of the motion) of
Lotka--Volterra are indeed integrable; for integrable cases, the
construction of Lagrangians leads to new related equations, involving
symmetries (of the evolution vector $V$) which generate the kernel of the
symplectic $2$--form. The Euler--Lagrange equations that arise possess
several types of invariance, the simplest of which is time
reparameterization, which arises when the kernel of the symplectic $2$--form
is generated by the evolution vector $V$; finally, two examples (related to
known chaotic systems in the literature\cite{Arn80,{Van78}}) show that
Lagrangian descriptions for systems near chaotic behavior do exist and they
satisfy Shil'nikov conditions\cite{Arn81} on the existence of strange
attractors around the fixed points, though it is still uncertain whether a
homoclinic curve exists or not for such systems.

\section{Preview}
\label{secPreview}
Consider the autonomous equations of motion:
\begin{equation}
\label{motion3D}
 \dot{x}\u{}{a}(t) = V\u{}{a}[x\u{}{b}(t)] \quad ,\quad {\mathrm{a, b}}= 1,\, 2 , \,3\,,
\end{equation}
where $V$ defines a \textbf{flow}, velocity or evolution vector field in the
corresponding space, being a function (which is local in time) of the
\textbf{coordinates} $x\u{}{a}(t)$. As an example, the Lorenz equation is
defined by the vector
\begin{equation*}
  V[x,y,z]  \equiv \{V\u{}{x},V\u{}{y},V\u{}{z}\} = \{-\sigma \,x+\sigma \,y,
  \,r \,x - y - x\, z,\,x \,y - b \,z\}\,,
\end{equation*}
where $(\sigma,\,r,\,b) $ are three independent parameters.

\subsection{Lagrangian structures for three--dimensional systems}

In arbitrary dimensions, the problem of finding Lagrangian formulations or
{Action Principles} for the given equations of motion (\ref{motion3D}) leads
to covariant equations\cite{Hoj81} in terms of Lie derivatives of tensor
structures along the flux vector. In three dimensions, the equations are
very restricted: as it is an odd--dimensional system, the Euler--Lagrange
equations always have an invariance. For example, time reparameterization
invariance appears when the {kernel} of the symplectic $2$--form is
generated by the evolution vector field $V$.

From now on, unless explicitly stated, we will be dealing with objects which
are \textbf{time--independent}, in the sense that they only depend on time
implicitly, through their dependence on the coordinates $x\u{}{a}(t)\,.$

The equations of motion (\ref{motion3D}) are related to a
\textbf{Variational Principle} with \textbf{Action}
\begin{equation} \label{action}
 {S}[x\u{}{a}(t),t] =
\int_{t\d{}{o}}^{t_1} dt\,\bigg(L\d{}{a}\,\left( \dot{x}\u{}{a} - V\u{}{a}
\right)+ K\bigg)
\end{equation}
(we assume from here on a summation over repeated indices), where the
$1$--form $L[x\u{}{b}]$ and the zero-form $K[x\u{}{b}]$ satisfy the
following equation:
\begin{equation*}
   L\d{}{a,b} V\u{}{b} + L\d{}{b} V\u{}{b}{}\d{}{, a}  =
 K\d{}{ , a}\,,
\end{equation*}
with $K\d{}{ , a}\equiv \frac{\partial K}{\partial x\u{}{a}}\,.$

We rewrite the above equation in terms of invariant structures:
\begin{equation*}
 \Lie{V}{} L = d K\,,
\end{equation*}
where $\Lie{V}{}$ is the \textbf{Lie derivative} along the vector $V$, and
$d$ is the \textbf{exterior differential} (see\cite{Nak92} for a definition
of these operators).

We call the pair $(L;\,K)$, $1$--form $L$ and $0$--form $K$, a
\textbf{standard Lagrangian pair} for $V$ if $K \neq 0$. In the special case
$K = 0$ we call $L$ a \textbf{non--standard Lagrangian }for $V$: this case
allows for the construction of Poisson brackets and constants of the motion
(see section \ref{secApp}).

The \textbf{Euler--Lagrange} equations which come from the Action
(\ref{action}) are:
\begin{equation}\label{EL}
\Sigma\d{}{a b} \left( \dot{x}\u{}{b} - V\u{}{b}\right) = 0\,,
\end{equation}
where $\Sigma \equiv d L$ is the \textbf{symplectic $2$--form }or Lagrange
bracket whose components are:
\begin{equation*}
\Sigma\d{}{a b} =  L\d{}{b,a}- L\d{}{a,b}\,.
\end{equation*}

Notice that these Euler--Lagrange equations do not imply the original
equations of motion (\ref{motion3D}); instead they imply deformed or mixed
equations involving the \textbf{kernel} of the symplectic $2$--form. In
three dimensions, the symplectic $2$--form always has a kernel, and the
simplest thing to do is to ask the kernel to be proportional to the
evolution vector $V$ itself (for other types of kernel see section
\ref{subsecMore}):
\begin{equation*}
  \Sigma\d{}{a b} = \rho[x\u{}{d}] \epsilon\d{}{a b c} V\u{}{c}\,,
\end{equation*}
where $\epsilon\d{}{a b c}$ is the {Levi--Civita} $3$--form, a totally
antisymmetric tensorial density, with $\epsilon\d{}{1 2 3} = 1\,,$ and
$\rho[x\u{}{a}]$ is the \textbf{volume element}, a scalar density which
solves\cite{Hoj92}
\begin{equation}\label{density} (\rho V\u{}{a})\d{}{, a}  = 0\,,
\end{equation}
which is sufficient and necessary in order that $\Sigma$ be a closed
symplectic $2$--form for the evolution vector $V$. In terms of a {metric}
$g\d{}{a b}$, the volume element is $\rho = \sqrt{\mathrm{det}\, g\d{}{a
b}}$ and the above equation is equivalent to $\nabla\d{}{a} V\u{}{a} = 0$,
where the covariant derivative is taken with respect to the metric $g$.
Geometrically, the existence of such metric leads to the construction of an
\textbf{integral invariant}\cite{And96} (Liouville theorem) of the form
\[I = \int_M \rho[\u{x}{a}]\,d^3 x\,,\]
where $M$ is a region of the configuration space which is assumed to evolve
along the flux $V$. The volume element is not unique: in fact, given two
volume elements $\rho, \rho'$, their quotient $C =\rho/\rho'$ is a constant
of the motion.

As a consequence of the fact that the kernel of the symplectic $2$--form is
generated by the evolution vector, our Action Principles for
three--dimensional systems will be purely geometrical: they will give only
the curve of the motion, leaving the way local clocks run undetermined. The
Euler--Lagrange equations (\ref{EL}) are thus $\Sigma\d{}{a b}
\dot{x}\u{}{b}  = 0\,,$ which are equivalent to $\frac{d {x}\u{}{a}}{d s} =
V\u{}{a} \,,$ where $s$ is an arbitrary parameter.

The standard Lagrangian pair ($L$;\,$K$) is constructed (locally) by a line
integral:
\begin{equation}\label{preLagr}
  \begin{array}{rcl}
    K[x\u{}{b}]  & = & L\d{}{a}[x\u{}{b}]\, V\u{}{a}[x\u{}{b}] \,,\\
    L\d{}{a}[x\u{}{b}] & = & \int_{0}^{1}\,\Sigma\d{}{a b}[x\u{}{c}(s)]\,\frac{dx\u{}{b}(s)}{ds} \,s\,ds + R\d{}{,a}[x\u{}{b}]\,,
  \end{array}
\end{equation}
where $R$ is an arbitrary $0$--form and the path in the line integral is
parameterized by
\[x\u{}{a}(s) = x\d{}{o}\u{}{a} + s\,(x\u{}{a}-x\d{}{o}\u{}{a})\,,\quad s \in [0,1]\,,\]
and is such that the integrand is equal to zero at $s=0$ and well behaved
along the path. The Action for the system shows clearly time
reparameterization invariance:
\begin{equation*}
 {S}[x\u{}{a}(t)] =
\int_{t\d{}{o}}^{t_1} L\d{}{a}[x\u{}{b}(t)]\, \dot{x}\u{}{a}(t)\,dt\,.
\end{equation*}

It is important to mention here the closeness of the equation for the volume
element with that for the \textbf{Darboux polynomial}. In the usual
terminology\cite{Cai00}, two polynomials in the coordinates $f(x\u{}{a})$
(the Darboux polynomial) and  $Q(x\u{}{a})$ (the \textbf{cofactor}) are to
be found in such a way that
\[f\d{}{,a} V\u{}{a} = Q\,f\,,\]
therefore we may get a solution for the volume element, equation
(\ref{density}) from a Darboux polynomial when the cofactor is a multiple of
$V\u{}{a}\d{}{,a}$. There are usually other solutions for the volume element
such that the resulting expression for $f$ is non--polynomial, as it happens
with the so--called exponential factors in Ref.\cite{Cai00a}.

Anyway, this closeness may be used directly to construct Action Principles
for every equation for which a Darboux polynomial or ``Darboux function" is
known, as it happens in the Lorenz equation, the R\"ossler equation, the
Lotka--Volterra equation and many others.

\section{Applications}
\label{secApp} In the following applications, we assume that the evolution
vector $V$ possesses a non--standard Lagrangian $L$ (this fixes $R$ in
equation (\ref{preLagr})).
\subsection{Method to construct a constant of the motion starting from a
known one} We may use a known constant of the motion to construct the other
explicitly. We will depict the method, omitting the details for simplicity.

Assume that the evolution vector $V$ possesses: (a) A non--standard
Lagrangian $L$, along with its associated volume element $\rho\,$; (b) a
time--independent constant of the motion $H[x\u{}{a}]\,.$ Consider the
vector defined by the components
\begin{equation*}
\eta\u{}{a} = \frac{1}{\rho}\,\epsilon\u{}{a b c} H\d{}{,b} L\d{}{c}\,.
\end{equation*}

It follows that $\eta$ is a symmetry for $V$, and that $\eta$ is
proportional to $V$. Therefore we obtain
\[ \eta\u{}{a} = I \,V\u{}{a}\,,\]
where $I$ is a constant of the motion for $V$, which may or may not be
independent of $H$. We will show that the system is completely
\textbf{integrable} (i.e., it possesses two independent, time--independent
constants of the motion) in any case:
\begin{enumerate}
\item{if $I=0$, then $D\,H\d{}{,a} = L\d{}{a}$ where $D$ is an $H$--independent constant of the motion, because, otherwise, $d L$ would vanish,}
\item{if $I \neq 0$ is independent of $H$, it is direct and}
\item{if $I \neq 0$ is a function of $H$, $I=I(H)$, define the $0$--form $P(H)$
s.t. $\frac{d P}{d H} = \frac{1}{I(H)}\,.$ Define next the $1$--form $U =
{\mathrm{e}}^{-P}\,L\,.$ It follows that $d U = 0$, therefore the $0$--form
$C$ defined (locally) by
\[d C = {\mathrm{e}}^{-P}\,L\]
is an $H$--independent constant of the motion for $V$.}
\end{enumerate}

\subsection{Hamiltonian theories}
Using the constants of the motion obtained in the last subsection, it is
easy to construct all possible Hamiltonian theories for the evolution vector
$V$. For example, for cases (ii) and (iii) above we may write
\[V\u{}{a} = J\u{}{a b} H\d{}{,b}\,,\]
where $J\u{}{a b} = {(I\, \rho)^{-1}}\,\epsilon\u{}{a b c} L\d{}{c}\,$ is
the \textbf{Poisson Bracket}. The \textbf{Casimir}, though always
computable, is easy to guess only in case (iii): it is the above $0$--form
$C$, and it solves $J\u{}{a b}C\d{}{,b} = 0\,.$ The \textbf{Hamiltonian} is
$H$ modulo a function of $C$; its final selection relies on stability
conditions (see Ref.\cite{Mor98}).

\subsection{More Lagrangian Theories}\label{subsecMore}
If we have two independent constants of the motion for system
(\ref{motion3D}), then another time--dependent constant of the motion may be
constructed (in principle) by integration of any component of the evolution
vector. If $H^1[x,y,z],\, H^2[x,y,z]$ are two constants, assume we may solve
for $y$ and $z$, getting $y = y[x,H^1,H^2],\, z = z[x,H^1,H^2]$. Thus the
first components of the equations of motion, $\dot{x} = V^1[x,y,z]$, are
transformed into $\dot{x} = F[x,H^1,H^2]$ and thus
\begin{equation*}
  \widetilde{{H}}^3[x,y,z,t] = H^3[x,y,z] - t\,
\end{equation*}
is a time--dependent constant of the motion, where $H^3[x,y,z] =\left.
\left(\int^x\, \left(F[x,H^1,H^2]\right)^{-1}\,dx \right)\right|_{H^j =
H^j[x,y,z]}$\,.

Now that we have three constants of the motion, we assert that the following
is the generic form for any non--standard Lagrangian $1$--form:\cite{Hoj81}
\[L = C^1\,d C^2\,,\]
where $C^j = C^j[H^1,H^2,H^3-t]\,,\quad j=1,2\,.$

In the case that any $C^j$ depends on $H^3$, the associated kernel of the
symplectic $2$--form is not anymore proportional to $V$: indeed it is
proportional to a (probably time--dependent) symmetry $\eta$ for $V$, which
may be calculated explicitly. We get, as a result, new Euler--Lagrange
equations which mix the vector $V$ with $\eta$. Examples with
time--independent symmetries $\eta$ will be shown in section
\ref{subsecABC}.

\subsection{Symmetries}
Finally, assuming that the $0$--forms $H^1,H^2,H^3\,$ from the last
subsection are known, then an Abelian algebra of vectors $\mathcal{A} =
\{\eta_1,\,\eta_2,\,\eta_3 \equiv V\}$ is constructed by taking the dual
vectors of the $1$--forms $d H^1,d H^2,d H^3$, i.e., by solving for the
vectors:
\begin{equation*}
  \Lie{\eta_j}{H^k} = \delta^k_j\,,\quad j,\,k = 1,\,2,\,3\,.
\end{equation*}

These commuting vectors are symmetries of each other by definition, and they
are used to generate the kernel of the most general symplectic $2$--form for
the evolution vector $V$ (see section \ref{subsecABC}).

\section{Examples: The Lotka--Volterra equation} \label{secExamples}
 This equation,
restricted to three dimensions, reads in its general form,
\begin{equation*}
\begin{array}{ccccl}
\dot{x}&=&  V\u{}{x}[x,y,z] & = & x\,(a_1 + b_{11}\,x + b_{12}\,y + b_{13}\,z) \\
 \dot{y}&=&  V\u{}{y}[x,y,z] & = & y\,(a_2 + b_{21}\,x + b_{22}\,y + b_{23}\,z)  \\
\dot{z}&=&  V\u{}{z}[x,y,z] & = & z\,(a_3 + b_{31}\,x + b_{32}\,y +
b_{33}\,z) \,,
\end{array}
\end{equation*}
where $a_{i},\,b_{i j}$ are constant parameters. We look for a volume
element of the form
\begin{equation}\label{densLV}
\rho[x,y,z] = x^u\,y^v\,z^w\,,
\end{equation}
where $u,v,w$ are constants that depend on the above parameters.

For some values of the parameters, non--standard Lagrangian $1$--forms may
be found: their existence allow to define Poisson brackets, and show
explicit integrability. In other cases, only standard Lagrangian pairs are
obtained, though the question if these Lagrangians could be made
non--standard by an addition of a closed $1$--form is open. The examples
suggest that, for chaotic or near chaotic systems, the answer to this
question is negative.

\subsection{Case $b_{ii} = 0\,,\quad i = 1,\,2,\,3\,$ (no Verhulst terms)}

The evolution vector is, after rescaling,
\begin{equation}\label{LV1}
V[x,y,z]  = \{ x\,(\lambda + C\,y + z),
            \, y\,(\mu + A\,z + x) ,
            \,z\,(\nu + B\,x + y)\}\,,
\end{equation}
where $A,\,B,\,C,\,\lambda,\,\mu$ and $\nu$ are constant parameters.
Constants of the motion for this system exist for a subset of the parameter
space.\cite{Str88,Cai00,Cai00a,{Mou01}} For arbitrary values of the
parameters, however, this system is not integrable, but as we will see, a
Lagrangian description always exists. Consider the volume element
\[\rho[x\u{}{a}] = (x \,y \,z)^{-1}\,.\]

It is easy to see that $(\rho V\u{}{a})\d{}{, a} = 0\,,$ and the Lagrangian
$1$--form is obtained directly:
\begin{equation}\label{StdLagr}
  L\d{}{a}[x\u{}{b}] =  (x\,y\,z)^{-1}\,\epsilon\d{}{a b
c}\,(x\u{}{b}\,V\u{}{c}[x\u{}{d}]+\eta\u{}{b}\,W\u{}{c}[x\u{}{d}]) \,,
\end{equation}
where $\eta\u{}{a} = M\u{}{a}\d{}{b} x\u{}{b}\,,$ $M\u{}{a}\d{}{b} =
\mathrm{diag}(\lambda,\,\mu,\,\nu)\,,$ and $W\u{}{a} = x\u{}{a}\,\ln
x\u{}{a}\,.$

Notice that the above volume element is singular at the planes $x=0$, $y=0$,
$z=0$, reflecting the fact that these are invariant planes.

\subsection{$ABC$ system}\label{subsecABC}
 When $\lambda = \mu =\nu =0\,,$ the
system defined by the flow (\ref{LV1}) is called ``$ABC$ system". In this
case, the Lagrangian (\ref{StdLagr}) is non--standard, and reduces to
\begin{equation*}
L\d{}{a}[x\u{}{b}] =  (x\,y\,z)^{-1}\,\epsilon\d{}{a b
c}\,x\u{}{b}\,V\u{}{c}[x\u{}{d}] \,.
\end{equation*}

The fact that this Lagrangian is non--standard leads us to conclude: if (for
given values of the parameters) there is a constant of the motion, then
another may be found using the theory of section \ref{secApp}. This may be
applied, for example, to all the $ABC$ systems studied in Ref.\cite{Mou01}
where the author finds one polynomial constant of the motion. In the next
examples, the method will be applied to some new cases found in
Ref.\cite{Cai00a}.

As a final aside, the Lagrangian may be used to find Poisson Brackets and
Hamiltonian theories, and to find new, related equations of motion from the
construction of a wide class of Lagrangian theories, as it will be done in
the last example of this subsection.

\begin{enumerate}
\item{Case $A=-1$,\, $B=1/2$,\, $C=0$.}

This case possesses a recently found\cite{Cai00a} constant of the motion:
$H^1 = x y^{-1} z^2 \exp{(-2(y+z)^2 (x y)^{-1})}$. We apply our Lagrangian
to construct another constant of the motion. In the present case, it turns
out
 that $U=(H^1)^{-1/2} L $ is a closed $1$--form, which therefore must be
the exterior derivative of some constant of the motion. We obtain, after
integration,
\begin{equation*}
  H^2[x,y,z]=(H^1)^{-1/2}\,x - 2  \int_0^{(y+z)/\sqrt{x
  y}}\exp{(q^2)}\,dq\,,
\end{equation*}
and thus this system is integrable.

\item{Case $A=-1/(C+1)$,\, $B=2$.}

This is a new case also found in Ref.\cite{Cai00a}, with one constant of the
motion:
$H^1[x,y,z]=x^2|y|^{2(C+1)}|z|^{-2C}|2A^2\,x\,z-(y-A\,z)^2|^{C-1}\,.$
According to our theory, in this case we obtain that the $1$--form
$U=(H^1)^{-\frac{1}{2(1+C)}} L$ is closed. We integrate and get the other
constant of the motion:
\begin{equation*}
\hspace{-2cm}{ H^2[x,y,z]=(y-A\,z)\left(1-\Omega+x\,y\,\Omega^{A
  C}\,(1-\Omega)^{\frac{1}{2}+A}\,\frac{\Gamma(A)}{\Gamma(-AC)}\,\mathrm{B}_\Omega\left(-A C,\,\frac{1}{2}-A\right)\right)}\,,
\end{equation*}
where $\Omega = \frac{2\,A^2\,x\,z}{(y-A\,z)^2}$ and $\mathrm{B}_\Omega(a,b)
\equiv \int_0^{\Omega}q^{a-1}\,(1-q)^{b-1}\,dq $ is the incomplete Beta
function.

\item{Case $ABC-1=0$,\, $B(A+1)+1=0$.}

Here a quadratic constant of the motion\cite{Str88,Cai00a} is known:
$H^1[x,y,z]=A^2(B\,x-z)^2-2\,A(B\,x+z)y+y^2\,.$ We obtain the other
constant:
\begin{equation*}
  \begin{array}{rcl}
  H^2[x,y,z]&=&|x|^{-1}\,|y\,z|^{-1-C}\times\left|\pm \sqrt{H^1} + A(B\,x-z)-y\right|^{1+2\,C}\\
             & & \times \left|(y+A\,z)x-(y-A\,z)C(\pm \sqrt{H^1}+y-A\,z)\right|\,.\
  \end{array}
\end{equation*}

The above three systems are, therefore, integrable and the construction of
many Hamiltonian as well as Lagrangian theories is possible. In the next
example we will take a known integrable case to show the construction of
Lagrangian theories which give new equations of motion, in terms of the
symmetries of the evolution vector field.

\item{Case $ABC+1=0$,\, $C=1$.}

Here, the integrability is guaranteed by the first condition\cite{Cai00a}.
We use the restriction $C=1$ here for simplicity only. Known constants are
$H^1[x,y,z]=-x + y -A\,z$ and $H^2[x,y,z]=x\,y^B\,z^{-1}\,.$ Now we take the
$y$--component of the equation of motion (\ref{motion3D}), to obtain the
third (time--dependent) constant of the motion:
\begin{equation*}
  \widetilde{{H}}^3[x,y,z,t] =   H^3[x,y,z] - t\,,
\end{equation*}
where
\begin{equation*}
 H^3[x,y,z] = -\frac{\ln\left( \frac{y}{x+A\,z}\right)}{H^1}\,.
\end{equation*}

The vectors $\eta_1,\eta_2,\eta_3$ dual to the $1$--forms $dH^1,dH^2,dH^3$
are commuting vectors, and they are found to be, after rearranging:
\begin{equation*}
\begin{array}{rcl}
{H^1}\,\eta_1[x,y,z]& = & \{x,y,z\}
  +H^3\,\eta_3[x,y,z]
  - {{B\,H^2}}\,\eta_2[x,y,z]\,,\\
\eta_2[x,y,z] & = & (x+A\,z)^{-1}\,y^{-B}\,z^2\, \{A,\,0,\,-1\}\,, \\
\eta_3[x,y,z]& = & V[x,y,z] = \{ x(y + z),\,y(x +A\,z),\,
  z(B\,x + y)\}\,. \
  \end{array}
\end{equation*}

It is clear from the definitions that any of the three equations of motion
(labelled by $j$) which the above vectors define,
\[\dot{x}\u{}{a} = \eta_j\u{}{a}[x\u{}{b}]\,,\quad j=1,2,3\,,\]
are completely integrable, with three constants of motion given by $C_j^{k}
= H^k - t\,\delta^k_j\,,\quad k=1,2,3.$

Now we turn to the Lagrangian descriptions for the evolution vector $\eta_3
=V$ (the same could be done for the other vectors). Among all the possible
examples, three special Action Principles are obtained with the Lagrangian
$1$--forms given by $L^1 =  H^2 dH^3$, $L^2 = - H^1 dH^3$ and $L^3 = H^1
dH^2\,.$ They are all non--standard Lagrangians for $V$, and the kernel of
the symplectic matrix $dL^j$ is easily found to be proportional to the
vector $\eta_j$. Therefore, the Euler--Lagrange equations coming from the
Lagrangian $L^j$ are
\begin{equation*}
  \dot{x}\u{}{a} = V\u{}{a} + \alpha\,\eta_j\u{}{a}\,,
\end{equation*}
where $\alpha$ is an arbitrary $0$--form. The case $j=3$ gives the usual
time reparameterization invariance, but the cases $j=1, 2$ are examples of
other kind of invariance; their Action Principles are easily computed from
equation (\ref{action}).
\end{enumerate}

\subsection{Lagrangians and chaos}

In the following examples we construct Lagrangians for Lotka--Volterra
systems which are close to known chaotic systems. The general results here
are: first, that a volume element of the form (\ref{densLV}), implies
$V\u{}{a}\d{}{,a} = 0$ at the finite \textbf{fixed point} (or singular
point) of $V$ which is not contained in any coordinate plane. This is
somewhat surprising, because the volume element is singular at the
coordinate planes, but its existence implies conditions on the vector field
at a point which is far from the planes.

Second, for a large subset of the parameter space, these systems allow for
Lagrangian descriptions along with Shil'nikov conditions on the existence of
a strange attractor of spiral type\cite{Arn81} in the vicinity of the
relevant fixed point, though no homoclinic curve has been found yet.

Third, the volume element becomes ill-defined (i.e., the powers $u,\,v,\,w$
in equation (\ref{densLV}) become infinite) when the parameter values are
such that some fixed points of the vector field degenerate into a ``fixed
line" (i.e., a line in the configuration space for which the evolution is
frozen) and the system gets a constant of the motion, which may be
calculated explicitly.

\begin{enumerate}
\item{A replicator--like equation.}

This example may be understood in the context of catalyst replication and
mutation:
\begin{equation}\label{repl}
\begin{array}{ccccl}
\dot{x}&=&  V\u{}{x}[x,y,z] & = & x\,(\frac{1}{2}(1-x)+\frac{1}{2}(1-y) + \frac{1}{10}(1-z)) \\
 \dot{y}&=&  V\u{}{y}[x,y,z] & = & y\,(-\frac{1}{2}(1-x)-\frac{1}{10}(1-y) + \frac{1}{10}(1-z)) \\
\dot{z}&=&  V\u{}{z}[x,y,z] & = & z\,(\lambda \,x + \mu(1-x)
+\frac{1}{10}(1-y)+\frac{1}{10}(1-z))\,,
\end{array}
\end{equation}
where $\lambda$ is a real parameter, and $\mu=-\frac{1}{6}-\lambda\,.$ The
volume element takes the form
\[ \rho =
{x^
     {-\frac{1 + 6\,\lambda }{6\,\lambda }}
     \,y^
     {-\frac{1 + 6\,\lambda }{3\,\lambda }}
     \,z^
     {\frac{1 - 2\,\lambda }
       {2\,\lambda }}}\,.\]

It is shown that $(\rho V\u{}{a})\d{}{, a} = 0\,.$ The Lagrangian $1$--form
is found to be
\begin{eqnarray}
\nonumber  \{L\d{}{x},L\d{}{y},L\d{}{z}\} = \rho &\{&\left( 1 + 5\,x - 3\,y
- 3\,z \right)
      \,y\,z + \left( -30 + 60\,x - 9\,y
      \right) \,y\,z\,\lambda ,\\
      & &\,0,\,
  3\,x\,y\,\left( -11 + 5\,x + 5\,y + z
      \right)  + 45\,x\,y^2\,\lambda \}\,.
 \
\end{eqnarray}

We note that the case $\lambda = 0$ and $\mu$ arbitrary is found in the
literature, displaying a one--parameter family of strange
attractors\cite{Arn80}. A bifurcation diagram in terms of $\mu$ may be found
in Ref.\cite{Phi00}. We mention it because our case intersects with the
latter at the point $\lambda = 0\,,\,\mu = -\frac{1}{6}\,,$ where the volume
element associated to the Lagrangian explodes. This point in the parameter
space also represents a degeneration of some fixed points of the vector
field, into the fixed line $\{ 1 + \frac{3\,s}{10}, 1 - \frac{s}{2},1 + s
\}$, $s \in \mathbb{R}$, and the following turns out to be a constant of the
motion for the system (\ref{repl}) :
\[C[x,y,z] = |x| \, |y|^2\,|z|^{-3}\,.\]

\item{A one predator--two prey system.}

We consider one of the most important models of predation\cite{Van78} using
the Lotka--Volterra equation, namely the equation
\begin{equation*}
  \begin{array}{rcl}
    \dot{x} = V\u{}{x}[x,y,z]& = & x(r - \frac{r}{K}\,x-\frac{r}{K}\,y - b\,z)\\
    \dot{y} = V\u{}{y}[x,y,z]& = & y(r - \frac{r}{K}\,\alpha\,x-\frac{r}{K}\,y - (b-\epsilon)\,z)\\
    \dot{z} = V\u{}{z}[x,y,z]& = & z(c\,b\,x + c\,(b-\epsilon)\,y - d)\,.\
  \end{array}
\end{equation*}
The values of the parameters allow to describe competitive superiority of
prey $x$ over prey $y$ (case $\alpha >1$), and predator--avoidance advantage
of prey $y$ over prey $x$ (case $0 \leq \epsilon \leq b$). Now, if we keep
the parameters arbitrary, we get the condition for the existence of the
volume element:
\begin{equation}\label{condition}
  d = \frac{c\,K\,\epsilon^2}{b\,(\alpha-1)}\,,
\end{equation}
and the volume element is
\begin{equation*}
  \rho[x,y,z]=\frac{1}{x\,y\,z}\left(x^{-\frac{\epsilon}{b}}\,y^{\frac{\epsilon}{b-\epsilon}}\,z^{\frac{r(\alpha-1)}{c
  K(b-\epsilon)}}\right)^{\frac{1}{\alpha-\alpha_{\rm o}}}\,,
\end{equation*}
where $\alpha_{\rm o} \equiv \frac{b}{b - \epsilon } - \frac{\epsilon
}{b}\,$.

In Ref.\cite{Van78}, the author finds a set of values of the parameters for
which the system develops spiral chaos.\cite{Arn81,{Ros76}} The values are
$\alpha =1.5,\, b=0.01,\,c=0.5,\,d=1,\,r=1,\,K=1000,$ and $\epsilon =
0.009\,.$

It is easy to check that there is no volume element of the form
(\ref{densLV}) for the above set of values, because condition
(\ref{condition}) is not met. We get that for the existence of the volume
element the only change is $d = 8.1$, which is much larger that the older
value, $d=1$. According to the model, this means that the predator ($z$
variable) has a larger mortality rate when there is a volume element of the
form (\ref{densLV}) than in the chaotic case.

On the other hand, a study of this system under condition (\ref{condition})
shows that, at the fixed point $(x\d{}{o},y\d{}{o},z\d{}{o}) \in
\mathbb{R}_- \times \mathbb{R}_+^2$, Shil'nikov criteria on the existence of
a strange attractor of spiral type\cite{Arn81} are met, for a large subset
of the parameter space. However, no homoclinic curve or chaotic behavior
have been found numerically yet.

The volume element becomes ill--defined when $\alpha = \alpha_{\rm o}\,$:
again, some fixed points of $V$ degenerate into a fixed line and the system
gets the following constant of the motion:
\begin{equation*}
  C[x,y,z] = |x|^{-\frac{d}{r\,\epsilon}(b-\epsilon)}\,|y|^{\frac{d}{r\,\epsilon}
  b}\,|z|\,.
\end{equation*}

\end{enumerate}

\section{Conclusions}

Lagrangian descriptions for three--dimensional systems are directly related
to the existence of determinants of metrics such that the covariant
divergence of the evolution vector is zero, which in turn implies that there
is an invariant volume element for the system. As a feature of
odd--dimensional systems, the Euler--Lagrange equations we get usually
possess time reparameterization invariance. A Variational Principle for
three--dimensional evolution equations may thus be useful for the study of
(quasi--)periodic orbits and long--time properties, or simply to test
numerical results.

In the case of the Lotka--Volterra equation, the very existence of the
volume element implies some condition on the vector field at the fixed point
which is relevant for the chaotic attractor. This condition is compatible
with Shil'nikov criteria for the existence of spiral chaos, but we have not
seen chaotic behavior numerically. Anyway, the condition does not imply at
all the integrability of the system. On the contrary, the examples show that
the volume element becomes ill--defined when constants of the motion appear
through degeneracy of fixed points of the flow into fixed lines.

Recently found quasi--integrable systems (e.g., $ABC$ systems with one
constant of the motion) are shown to be integrable using a newly devised
method to find a constant of the motion starting from a known one. Finally,
the Lagrangian viewpoint for integrable systems leads to Euler--Lagrange
equations with several kinds of invariance (including time
reparameterization), according to the kernel of the associated symplectic
$2$--form.

\ack

 One of us (M.B.) is deeply grateful to Fundaci\'on Andes (Grant
for Doctoral Studies) and Conicyt (Thesis Completion Scholarship), for
financial support.

\end{document}